\documentclass[prd,twocolumn,showpacs,showkeys]{revtex4}
\usepackage{epsfig}

\begin{document}

\title{Neutrino CP violating parameters from
  nontrivial quark-lepton correlation: a S3 x GUT model}

\author{Marco Picariello}
\email{Marco.Picariello@cern.ch}
\affiliation{
Dipartimento di Fisica, Universit\`a di Lecce and INFN-Lecce\\
Via Arnesano, ex Collegio Fiorini, I-73100 Lecce, Italia\\
 and\\
Centro de F\'{\i}sica Te\'{o}rica das Part\'{\i}culas (CFTP)\\
Departamento de F\'{\i}sica, Lisboa\\
Instituto Superior T\'{e}cnico\\
Av. Rovisco Pais, P-1049-001 Lisboa, Portugal
}

\begin{abstract}
We investigate the prediction on the lepton phases in theories with a non
trivial correlation between quark (CKM) and lepton (PMNS) mixing matrices.
We show that the actual evidence, under the only assumption that the correlation matrix
$V^M$ product of $CKM$ and $PMNS$ has a zero in the entry $(1,3)$, gives us a prediction
for the three CP-violating invariants $J$, $S_1$, and $S_2$.
A better determination of the lepton mixing angles will give stronger prediction for the CP-violating invariants in the lepton sector.
These will be tested in the next generation experiments.
To clarify how our prediction works, we show how a model based on a
Grand Unified Theory and the permutation flavor symmetry $S_3$
predicts $V^M_{13}=0$.

\pacs{14.60.Pq, 14.60.Lm, 96.40.Tv.}
\keywords{Neutrino physics, quark lepton complementarity, Grand Unified Theories, Flavor symmetries}

\end{abstract}

\maketitle

\section{Introduction}
After the recent experimental evidences about neutrino physics
\cite{SNO}%
-\cite{MINOS}
 we know very well almost all the parameters
in the quark \cite{Charles:2004jd}
and lepton \cite{Fogli:2005gs}%
-\cite{McDonald:2006qf}
sectors.
 We measured all the quark and charged lepton masses,
and the value of the difference between the square of the neutrino masses
$\delta m_{12}^2=m_1^2-m_2^2$ and $\delta m_{23}^2=|m_3^2-m_2^2|$. 
We also know the value of the quark mixing angles and phases, and the two mixing angles
$\theta_{12}$ and $\theta_{23}$ in the lepton sector.
The challenger for the next future
\cite{Petcov:2006gy}%
-\cite{Broggini:2006rm}
 will be to determine the sign of $\delta m^2_{23}$
(i.e. the hierarchy in the neutrino sector), the absolute scale of the neutrino masses,
and the value of the 3rd lepton mixing angle $\theta_{13}$
(in particular if is it zero or not).
Finally, if $\theta_{13}$ is not too small, there is a hope to measure the CP
violating phases.

Despite the fact that we have all these experimental evidences, still a theory of
flavor is missing. The main point is related to the fact that there is a hierarchy
in the mixing angles and masses. In particular the quark mixing angles
are in general much smaller then the corresponding lepton mixing angles.
The small quark mixing angles can be explained with continuous flavor symmetries
(see for instance \cite{Barbieri:1999km}),
while the structure of the lepton mixing matrix seems to be better accommodated 
in models with discrete symmetries
(see for example \cite{Altarelli:2006ri}%
-\cite{Caravaglios:2005gw}).
Recently, the disparity that nature indicates between quark and lepton mixing
angles has been viewed in terms of a 'quark-lepton complementarity' (QLC)
\cite{Minakata:2004xt,Ferrandis:2004mq} which can be expressed in the relations
\begin{equation}\label{eq:QLCnaive}
\theta_{12}^{PMNS}+\theta_{12}^{CKM}\simeq 45^\circ\,;
\quad\quad
 \theta_{23}^{PMNS}+\theta_{23}^{CKM}\simeq 45^\circ\,.
\end{equation}
These relations are related to the parametrization used for the CKM and PMNS mixing
matrix.
From a more general point of view, we can define a correlation matrix $V^M$
as the product of the PMNS  \cite{Pontecorvo:1967fh,Maki:1962mu} and
 CKM \cite{Cabibbo:1963yz,Kobayashi:1973fv} mixing matrices,
\begin{eqnarray}\label{eq:fundO}
V^M=U^{CKM}\, U^{PMNS}\,.
\end{eqnarray}
 
A lot of efforts have been done to obtain the
{\em most favorite} pattern for the matrix
$V^M$~\cite{Chauhan:2006im}%
-\cite{Antusch:2005ca}.
The naive QLC relations in eq. (\ref{eq:QLCnaive}) seems to
implies $V^M$ to be BiMaximal, i.e. in the standard parametrization
it contains two maximal mixing angle ,
and a third angle to be zero.
A BiMaximal $V_M$ implies a nice, in the sense that it is testable at
the near future neutrino experiments, prediction on the $\theta_{13}^{PMNS}$
mixing angle \cite{Minakata:2004xt} and the CP violating parameters for the
lepton sector \cite{Hochmuth:2006xn}.
At first order approximation however $V^M$ BiMaximal seems not to be
compatible with the experiments \cite{Xing:2005ur}.
From our previous work \cite{Chauhan:2006im} we learn that
$V^M$ BiMaximal, although it is not ruled out
by the experiments, is excluded at 90\% CL in non SUSY models,
or in SUSY models with $\tan \beta<40$ where the RGE correction are
negligible \cite{Kang:2005as}%
-\cite{Antusch:2005gp}.

Despite the fact that the correlation matrix $V^M$ cannot be BiMaximal, still
there are very nice phenomenological consequences from a non trivial $V^M$.
All of these consequences seems to be related to the fact that 
experimental evidences tell us \cite{Chauhan:2006im} that $V^M_{13}=0$
is in good agreement with the experimental data, and is even the
preferred value.
From a theoretical point of view we like very much the fact that
experimental data tell us that $V^M_{13}=0$.
First of all the theoretical ingredient of the quark-lepton complementarity
that gives phenomenological predictions is $V^M_{13}=0$.
In fact $V^M_{13}=0$, without any other assumptions on the full $V_M$ matrix
with the exception of the compatibility with the actual experimental evidences,
implies a testable prediction for the undetermined lepton mixing angle 
$$\theta^{PMNS}_{13}=(9^{+1}_{-2})^\circ\quad(\mbox{see [65]})$$
Second because $V^M_{13}=0$ can be achieved in a natural way in theoretical
models by imposing some symmetry.

In this work first we analyze the consequences of $V^M_{13}=0$
on the CP violating invariants for the lepton sector.
Then we show how a toy model, based on $S_3$ flavor permutation 
symmetry \cite{Caravaglios:2006aq} and GUT gives $V^M_{13}=0$.

We use a Monte Carlo simulation with two-sided Gaussian distributions
around the mean values of the observables to extract the $J$, $S_1$,
and $S_2$ invariants.
The input information on $\theta_{13}^{PMNS}$ is taken from the analysis of
ref.~\cite{Chauhan:2006im} which uses quark lepton correlation ($V^M_{13}=0$),
and neutrino and quark data.
We obtain that the main ingredient to obtain a prediction on the CP
violating invariants for the lepton sector is the constraint
$V^M_{13}=0$.

From the theoretical point of view we show that, if we do not take care
about a required fine-tuning in obtaining the masses,
it is possible to construct a toy model \cite{Caravaglios:2006aq},
based on $S_3$ flavor permutation symmetry
and GUT, that gives us $V^M_{13}=0$.
In Grand Unified Theories (GUT), as long as quarks and leptons are
inserted in the same representation of the gauge group, we need to include
in the definition of $V^M$ non trivial phases between the $CKM$ and $PMNS$
mixing matrices.
In our model $V^M$ is defined by
\begin{eqnarray}\label{eq:fund}
V^M=U^{CKM}\, \Omega\, U^{PMNS}\,.
\end{eqnarray}
where $\Omega$ is a diagonal matrix
$\Omega={\rm diag} (e^{{\rm i} \omega_i})$ and the
three phases $\omega_i$ are free parameters (in the sense that they are
not determined by present experimental evidences).  
The matrix $V_M$ is related to the Dirac and Majorana neutrino mass matrix.
For this reason its form is given by the symmetries of the model.

The paper is organized as follows: in section~{\bf\ref{sec:CP}} we
introduce our notation and the parameterization for CKM and PMNS mixing
matrices. With the aid of a Monte Carlo simulation,
we study the numerical correlations of the lepton
CP violating phases $J$, $S_1$, and $S_2$ with respect to the
mixing angle $\theta_{12}^{PMNS}$.
In section~{\bf\ref{sec:Model}} we clarify the relation between
the correlation matrix $V^M$ and the neutrino mass matrices
in GUT model, with a see-saw of type I.
After that we show how a $S_3$ flavor symmetry can give us
$V^M_{13}=0$.
Finally in section~{\bf\ref{sec:End}} we present a summary and our conclusions.

\section{CP violating invariants in the lepton sector}\label{sec:CP}
As usually, we parameterize the lepton mixing matrix as
\begin{eqnarray}\label{PMNS}
U^{PMNS}= U^{23} \Phi U^{13} \Phi^\dagger U^{12} \Phi^m
\end{eqnarray}
where $\Phi$ is a diagonal matrix with elements
$\{1, 1, e^{i \phi}\}$, and $\phi$ is the Dirac CP violating phase.
$\Phi^m$ contain the Majorana phases and is a diagonal matrix with
elements given by $\{e^{i \phi_1}, e^{i \phi_2}, 1\}$, and $U^{ij}$
are rotation in the $(i,j)$ plan.
There are two kind of invariants parameterizing CP violating effect.
The Jarlskog invariant [28] $J$ that parametrizes the effects related 
to the Dirac phase,
and the two invariants $S_1$ and $S_2$ that parametrize the effects related
to the Majorana phases.
The $J$ invariant describes all CP breaking
observables in neutrino oscillations [29].
It is the equivalent of the Jarlskog invariant in the quark sector.
It is given by
\begin{equation}\label{eq:JU}
J=Im\{U_{\nu_e \nu_1}U_{\nu_\mu \nu_2}U_{\nu_e \nu_2}^* U_{\nu_\mu \nu_1}^*\}\,.
\end{equation}
In the parametrization of eq. (\ref{PMNS}) one has
\begin{eqnarray}\label{eq:J}
J&=&\frac{1}{8}\sin2\theta_{12}\ \sin2\theta_{23}\ \sin2\theta_{13}\ \cos\theta_{13}\ \sin\phi\,.
\end{eqnarray}
Then we have the two invariants $S_1$ and $S_2$ 
that are related to the Majorana phases.
They are
\begin{eqnarray}\label{eq:SU}
S_1 &=& Im\{U_{\nu_e \nu_1} U_{\nu_e \nu_3}^*\}\\
S_2 &=& Im\{U_{\nu_e \nu_2} U_{\nu_e \nu_3}^*\}\nonumber
\end{eqnarray}
In the parametrization of eq. (\ref{PMNS}) we have
\begin{eqnarray}\label{eq:S}
S_1 &=& \frac{1}{2} \cos \theta_{12} \sin 2\theta_{13} \sin(\phi+\phi_1)\\
S_2 &=& \frac{1}{2} \sin \theta_{12} \sin 2\theta_{13} \sin(\phi+\phi_2)\nonumber
\end{eqnarray}
The two Majorana phases appear in $S_1$ and $S_2$ but not in $J$.

\subsection{Prediction for CP violating invariants}\label{sec:subCP}
In this section we investigate the consequences of a $V^M$
correlation matrix with a zero $(1,3)$
entry on the undetermined parameters $J$, $S_1$, and $S_2$.
We remember that $J$ is the Dirac invariant CP-violating phase,
and is the only one that can be observed in neutrino oscillations experiments.
As shown in a previous paper \cite{Chauhan:2006im}, the data favors a
vanishing $(1,3)$ entry in the correlation matrix $V^M$.
So in the whole analysis we fix $\sin^2 \theta_{13}^{V^M}=0$.
Moreover $\tan^2\theta_{12}^{V^M}$ and $\tan^2\theta_{23}^{V^M}$ are allowed
to vary respectively within the intervals $[0.3,1.0]$ and $[0.5,1.4]$.
We introduce the unitary Wolfenstein parameterization in terms of 
the variables $\lambda$ ,$A$, $\rho$, $\eta$ \cite{Wolfenstein:1983yz}
\begin{eqnarray}\label{CKM}
U^{CKM}=
 U^{23}\, \Phi\, U^{13}\,\Phi^\dagger\, U^{12}\,,
\end{eqnarray}
where one has the relations~\cite{Buras:1994ec}
\begin{eqnarray*}
\sin \theta_{12}^{CKM}&=&\lambda\\
\sin \theta_{23}^{CKM}&=&A\lambda^2\\
\sin \theta_{13}^{CKM}e^{-{\rm i} \delta^{CKM}}&=&A \lambda^3(\rho-{\rm i} \eta)
\end{eqnarray*}
to all orders in $\lambda$.
We use the updated values for the $CKM$ mixing matrix,
given at $95\%$CL by \cite{Charles:2004jd}
\begin{eqnarray}\label{bestfit}
\begin{tabular}{cc}
$\lambda = 0.2265^{+0.0040}_{-0.0041}$
\,,&
$A=0.801^{+0.066}_{-0.041}$
\,,\cr\cr
${\overline\eta} = 0.189^{+0.182}_{-0.114}$
\,,&
${\overline\rho} = 0.358^{+0.086}_{-0.085}$
\,,
\end{tabular}
\end{eqnarray}
with 
\begin{equation}
\rho + i\eta =
\frac{\sqrt{1-A^2\lambda^4}({\overline\rho}+i{\overline\eta})}{\sqrt{1-\lambda^2}
\left[1 - A^2\lambda^4({\overline\rho}+i{\overline\eta})\right]}\,.
\end{equation}
For the lepton mixing angle we impose \cite{Fogli:2005gs,Aliani:2003ns}
\begin{eqnarray}\label{bestfit2}
\sin^2\theta_{23}^{PMNS} &=& 0.44\times\left(1^{+0.41}_{-0.22}\right)\nonumber\\
\sin^2\theta_{12}^{PMNS} &=& 0.314\times\left(1^{+0.18}_{-0.15}\right)
\,,
\end{eqnarray}
and \cite{Chauhan:2006im}
\begin{eqnarray}
\theta_{13}^{PMNS} &=& \left(9^{+1}_{-2}\right)^\circ\,.
\end{eqnarray}
We allow the $U^{CKM}$ parameters to vary, with a two-sided Gaussian distribution,
within the experimental ranges given in eq. (\ref{bestfit}).
For the $\Omega$ phases in eq. (\ref{eq:fund}) we take flat distributions
in the interval $[0,2\pi]$.
We make Monte Carlo simulations for different values of $\theta_{12}^{V^M}$
and $\theta_{23}^{V^M}$ mixing angles, allowing $\tan^2\theta_{12}^{V^M}$
and $\tan^2\theta_{23}^{V^M}$ to vary respectively within their allowed intervals,
in consistency with the lepton and quark mixing angles.
From eq. (\ref{eq:J}), by using the fact that $\theta_{13}$ is small and that
$\theta_{23}$ is maximal, we get
$$
J\approx \frac{1}{8} \sin 2\theta_{12} \sin 2\theta_{13} \sin \phi
$$
This expression tells us that the $J$ parameter is within the range $|J|<0.042$.
However there is a non trivial correlation between $J$ and $\theta_{12}^{PMNS}$.
Because the $CKM$ is given by the experimental data,
and $V^M_{13}$ is fixed to be zero, the phase $\phi$ and the $\theta_{13}^{PMNS}$
angle are almost fixed as a function of $\theta_{12}^{PMNS}$.

\begin{figure}[ht]
\centering
{\epsfig{file=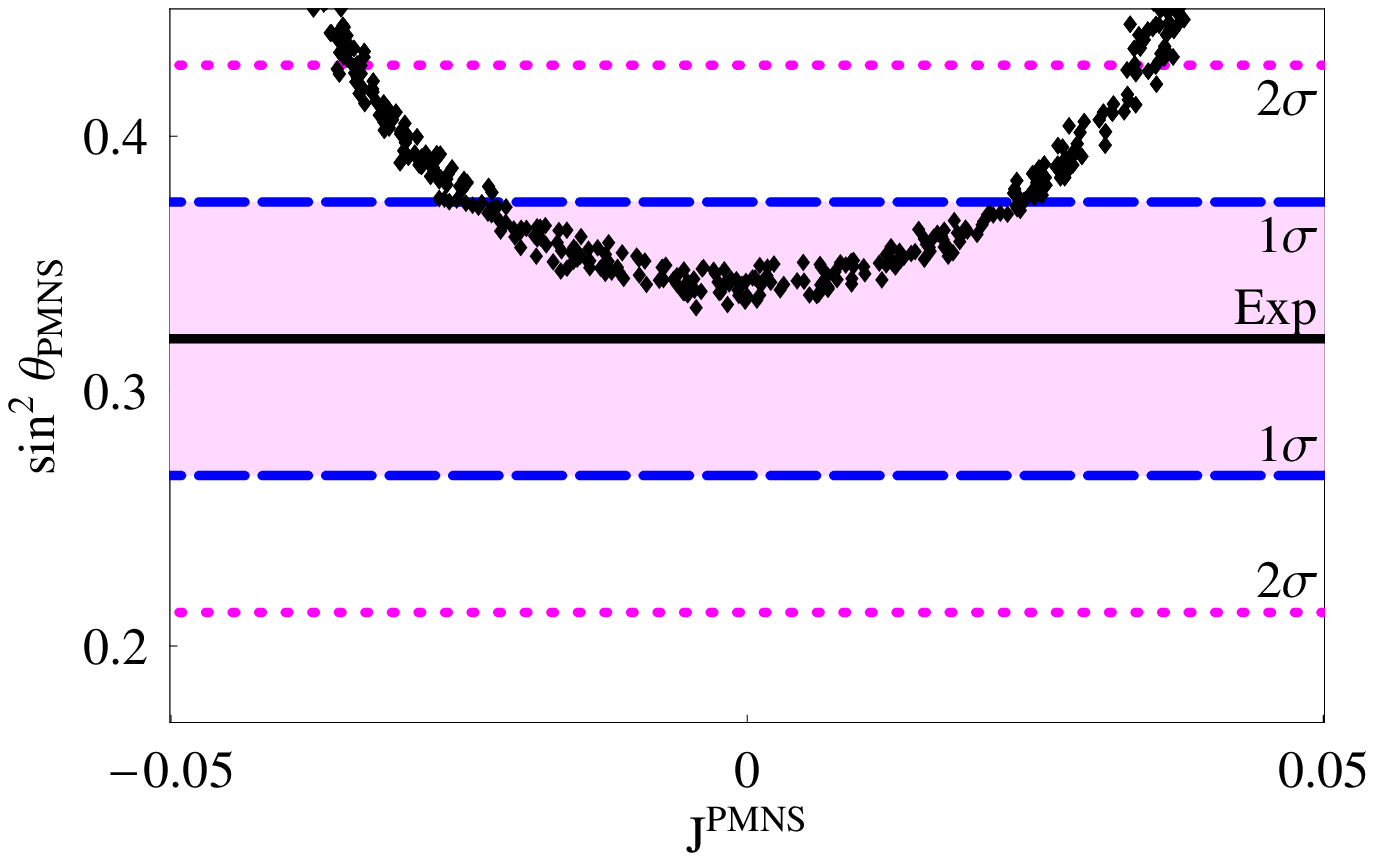,height=5.cm}}
\vskip -0.5truecm
\caption{The correlation between the Dirac CP violating 
parameter $J$ and $\sin^2\theta_{12}^{PMNS}$ for $V_M$
BiMaximal.
We also plot the experimental central value, the
$1\sigma$, and the $2\sigma$ for
$\sin^2 \theta_{12}^{PMNS}$.
}
\label{fig:f11}
%
\vskip 0.5truecm
\centering
{\epsfig{file=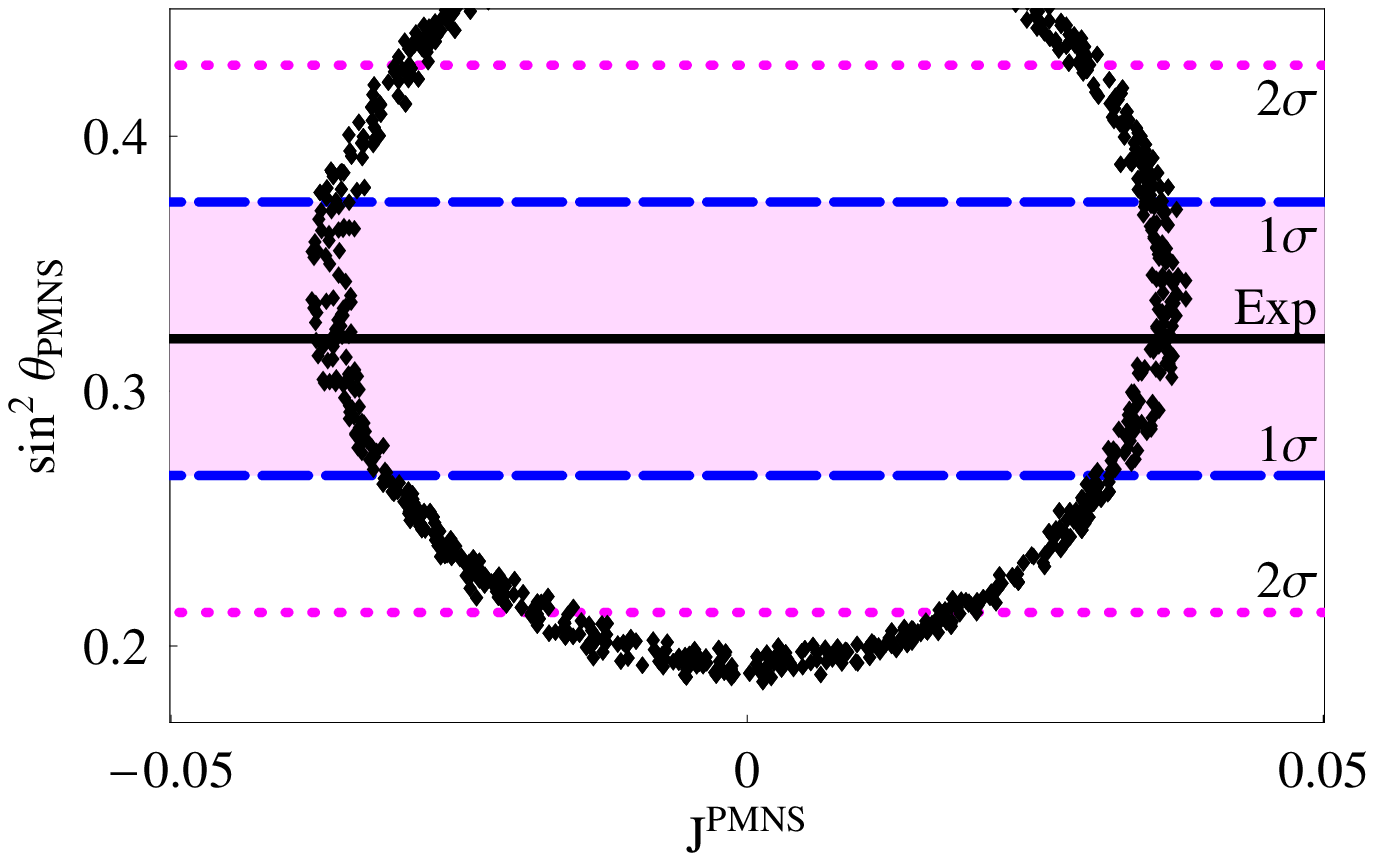,height=5.cm}}
\vskip -0.5truecm
\caption{The correlation between the Dirac CP violating 
parameter $J$ and $\sin^2\theta_{12}^{PMNS}$ for $V_M$ TriBiMaximal.
}
\label{fig:f12}
%
\vskip 0.5truecm
\centering
{\epsfig{file=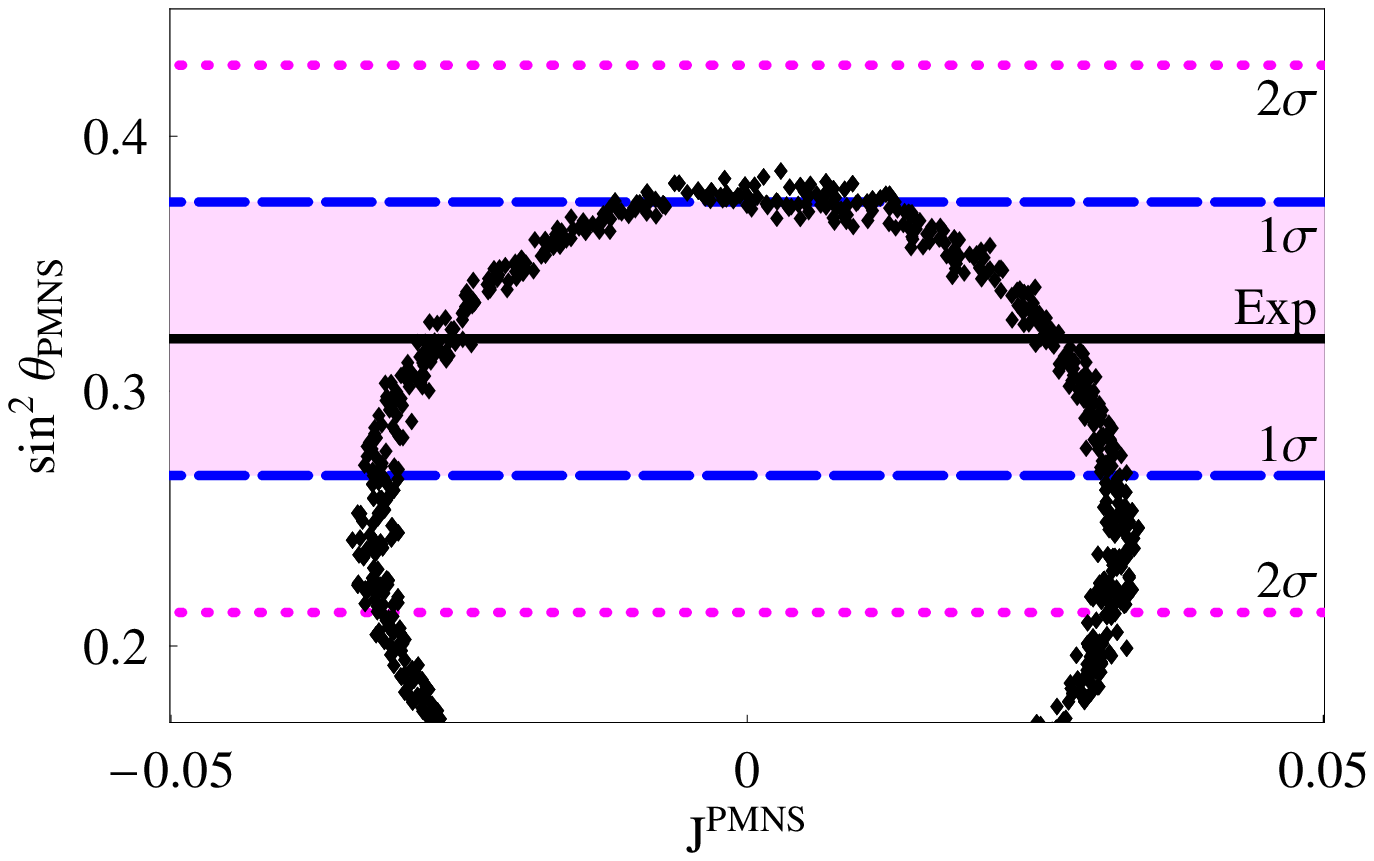,height=5.cm}}
\vskip -0.5truecm
\caption{The correlation between the Dirac CP violating 
parameter $J$ and $\sin^2\theta_{12}^{PMNS}$
for $V_M$ such that $\tan^2\theta_{12}^{V_M}=0.4$.
}
\label{fig:f13}
\end{figure}

In figs. \ref{fig:f11}-\ref{fig:f13}
we report the result of our simulation for $J$.
We plot the correlation between the $J$ invariant and $\sin^2\theta_{12}^{PMNS}$ for $V_M$ BiMaximal (fig. \ref{fig:f11}),
TriBimaximal (fig. \ref{fig:f12}),
and $V_M$ with $\tan^2\theta_{12}^{V_M}=0.4$ (fig. \ref{fig:f13}).
First of all, from fig. \ref{fig:f11}, we see that the {\em solar} mixing angle
$\theta_{12}^{PMNS}$ is constrained to have $\sin^2 \theta_{12}^{PMNS}>0.36$
for $V_M$ Bimaximal.
From figs. \ref{fig:f11}-\ref{fig:f13} we see the correlation between the
structure of $V_M$ and the CP violating invariant $J$. In particular,
for $V_M$ BiMaximal $J$ close to zero. For $V_M$ TriBiMaximal
$J$ is around $0.042$. Finally for $V_M$ such that
$\tan^2\theta_{12}^{V_M}=0.4$ we get that $J$ can be any value between
$-0.04$ and $0.04$. 
We also see that a better determination of the
$\sin^2\theta_{12}^{PMNS}$ could give a stronger
prediction for the $J$ invariant in the case of $V_M$ TriBimaximal.

\begin{figure}[ht]
\centering
{\epsfig{file=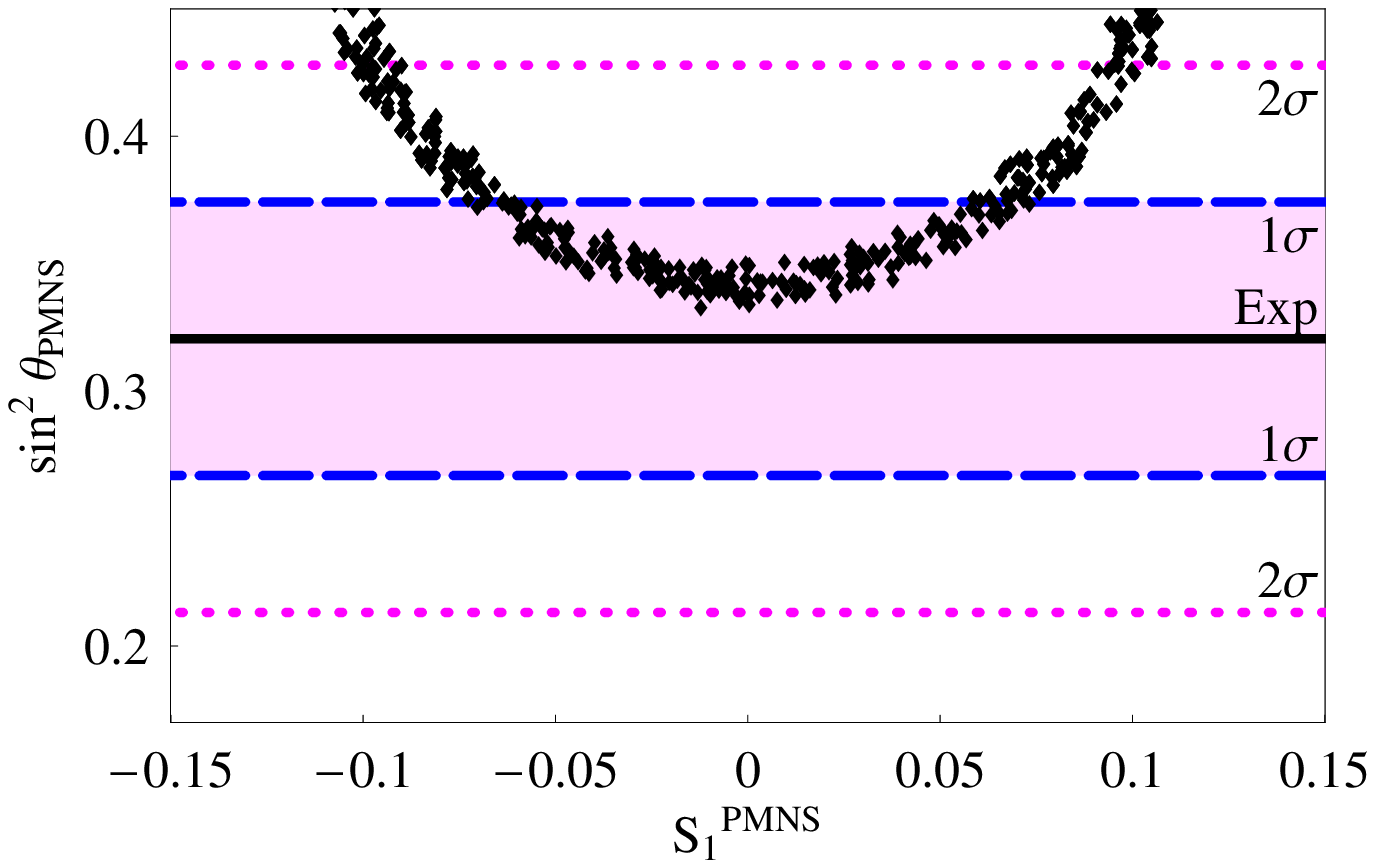,height=5.cm}}
\vskip -0.5truecm
\caption{Same as fig \ref{fig:f11} ($V_M$ BiMaximal)
for the correlation between the Majorana
 CP violating parameter $S_1$ and $\sin^2\theta_{12}^{PMNS}$.
}
\label{fig:f21}
%
\centering
\vskip 0.5truecm
{\epsfig{file=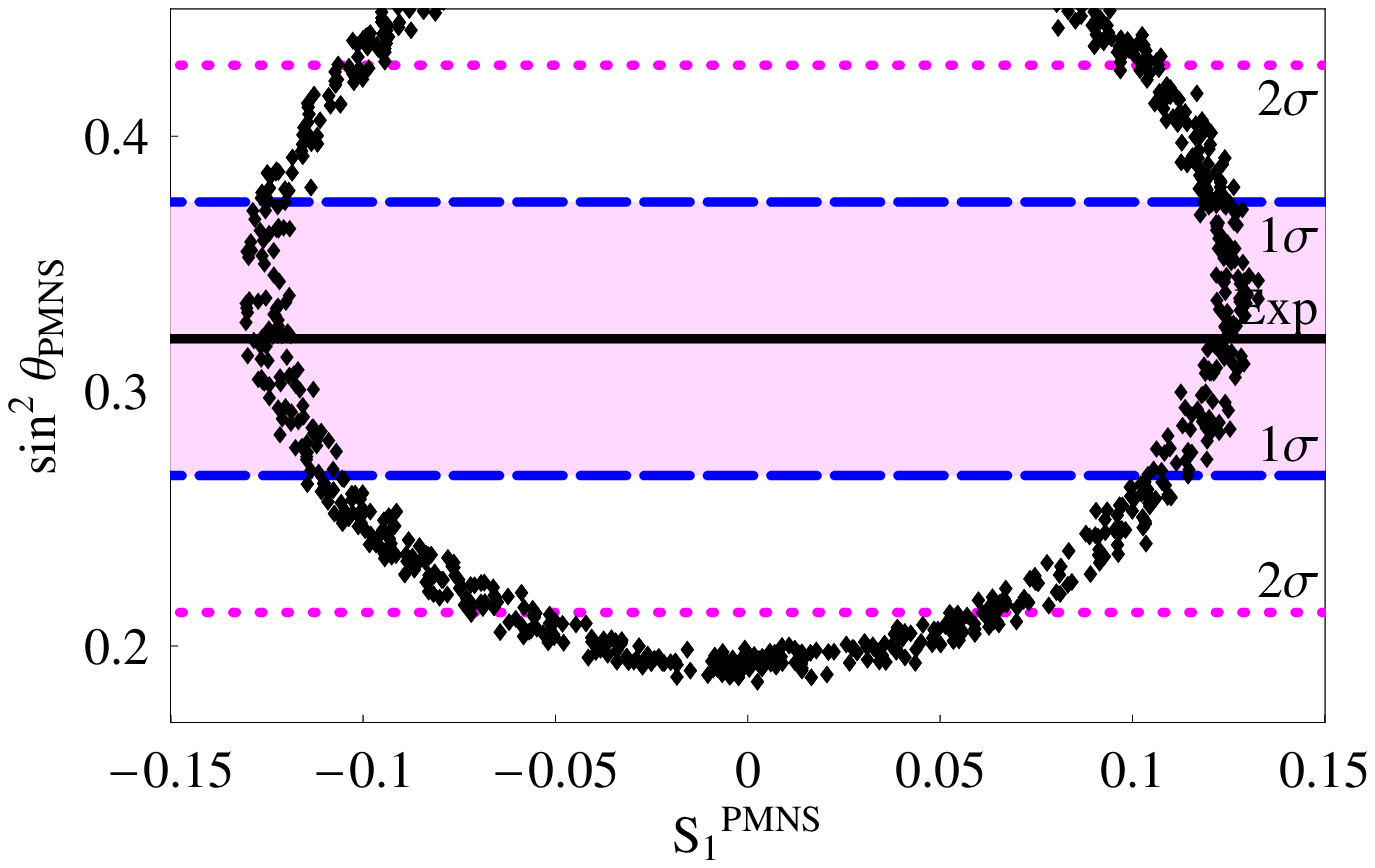,height=5.cm}}
\vskip -0.5truecm
\caption{Same as fig \ref{fig:f12} ($V_M$ TriBiMaximal)
for the correlation between the Majorana
 CP violating parameter $S_1$ and $\sin^2\theta_{12}^{PMNS}$.
}
\label{fig:f22}
%
\centering
\vskip 0.5truecm
{\epsfig{file=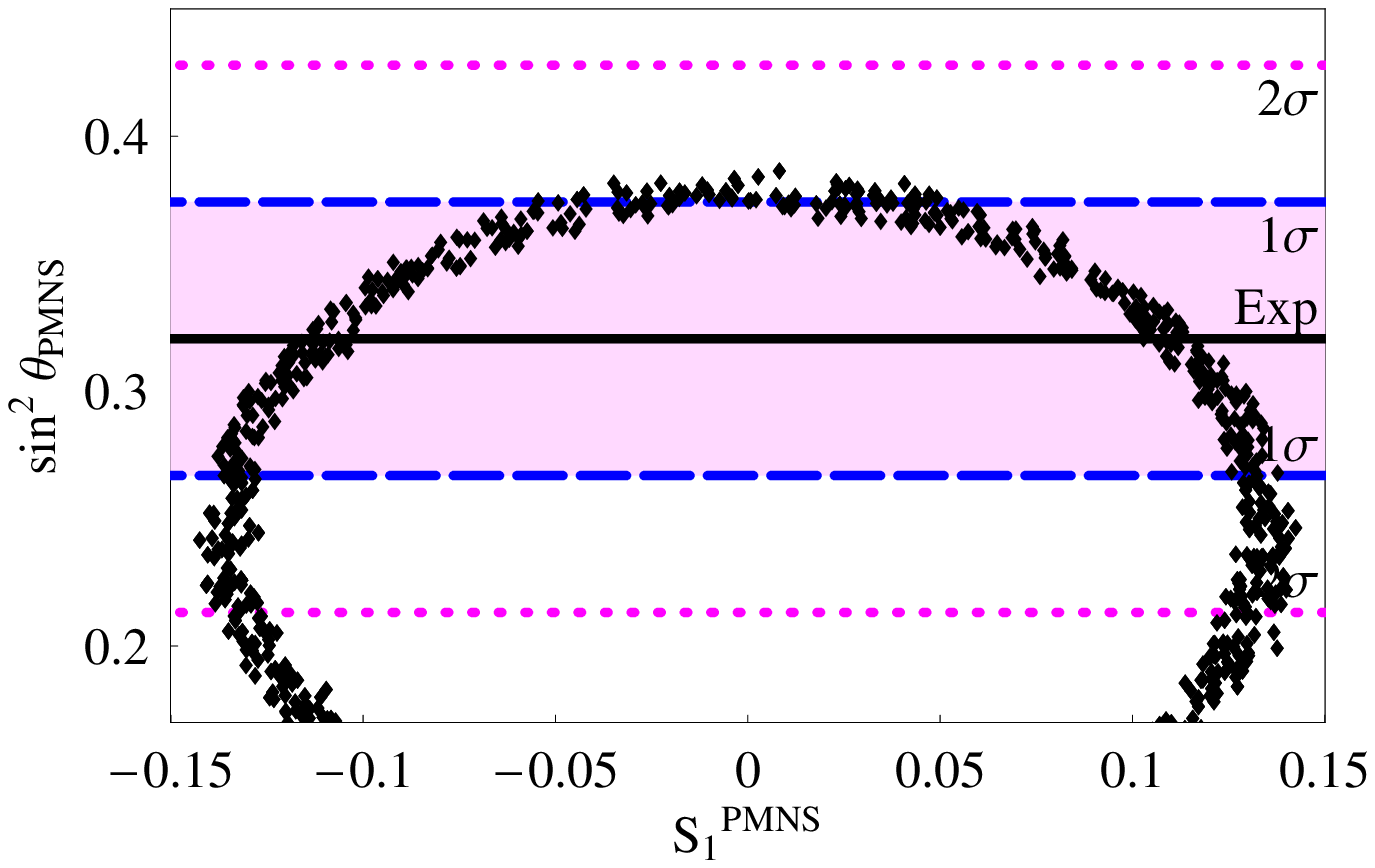,height=5.cm}}
\vskip -0.5truecm
\caption{Same as fig \ref{fig:f13} ($V_M$ such that
$\tan^2\theta_{12}^{V_M}=0.4$) for the correlation
between the Majorana
 CP violating parameter $S_1$ and $\sin^2\theta_{12}^{PMNS}$.
}
\label{fig:f23}
\end{figure}

In figs. \ref{fig:f21}-\ref{fig:f23} we report the result of our simulation
for $S_1$ ($S_2$ plots have similar shapes).
The expressions in eqs. (\ref{eq:S}) give us the range for these invariants:
\begin{eqnarray}
|S_1|<0.14&\quad\quad&|S_2|<0.11
\end{eqnarray}
We plot the correlations between the $S_1$ invariant 
with respect $\sin^2\theta_{12}^{PMNS}$ for $V_M$ BiMaximal
(fig. \ref{fig:f21}), TriBiMaximal (fig. \ref{fig:f22}),
and such that $\tan^2\theta_{12}^{V_M}=0.4$ (fig. \ref{fig:f23}).
From the figures we obtain that for $V_M$ BiMaximal the Majorana CP 
invariant $S_1$ is close to zero, for $V_M$ TriBiMaximal $S_1$ is around
$0.13$. Finally for $V_M$ such that $\tan^2\theta_{12}^{V_M}=0.4$
we obtain that $S_1$ can be any value between $-0.14$ and $0.14$.
Similar results hold for the other  Majorana CP violating
invariant $S_2$.
We see that also in this case a better determination of the $\theta_{12}^{PMNS}$ mixing angle will give a stronger constraint for the
$S_1$ (and $S_2$) invariant for $V_M$ TriBiMaximal.
As for $J$, these correlations of $S_1$ (and $S_2$) with respect to $\theta_{12}^{PMNS}$
are predictions of any theoretical GUT model that gives a relation of the type
$V^M=U^{CKM}\, \Omega\,  U^{PMNS}$ with $V^M_{13}=0$.
In the next section we will show how to construct an explicit model that predict
$V^M_{13}=0$.

\section{A toy model}\label{sec:Model}
In this section we will show how to construct a toy model that gives us the relation
$V^M=U^{CKM}\, \Omega\,  U^{PMNS}$ with $V^M_{13}=0$.
We will not take care of explicit values for the masses. To obtain them
we need an unwanted fine-tuning. However, as long as we reefer to
the mixing angles only, this model can be seen as a toy model explaining the relationship
between the CKM and the PMNS mixing matrix and the appearance of a zero $(1,3)$ entry 
in the quark-lepton correlation matrix $V^M$.

\subsection{$V^M$ in theories with see-saw of type I} 
Let's fix the notations in the lepton sector.
Let $Y_l$ be the Yukawa matrices for charged leptons.
It can be diagonalized by
\begin{eqnarray}
Y_l&=&U_l Y_l^\Delta V_l^\dagger
\end{eqnarray}
Let be $M_R$ the Majorana mass matrix for the right neutrino
and $M_{D}$ the Dirac mass matrix.
Under the assumption that the low energy neutrino masses are given by
the see-saw of Type I we have that the light neutrino mass matrix
is given by
\begin{equation}\label{eq:Mnu}
M_\nu = M_{D} \frac{1}{M_R} M_{D}^t\,.
\end{equation}
Let us introduce $U_0$ form the diagonalization of the Dirac mass
matrix
\begin{eqnarray}\label{eq:Dirac}
M_{D} &=& U_0 M_{D}^\Delta V_0^\dagger
\end{eqnarray}
then we define $V^M$ by the diagonalization of the light neutrino mass
\begin{eqnarray}
M_\nu &=& M_{D} \frac{1}{M_R} M_{D}^t = \nonumber\\\label{eq:light}
U_\nu M_\nu^\Delta U_\nu^t &=& U_0 V^M M_\nu^\Delta (V^M)^t U_0^t
\end{eqnarray}
Finally the lepton mixing matrix is
\begin{eqnarray}\label{eq:PMNS}
U^{PMNS} = U_l^\dagger U_\nu = U_l^\dagger U_0 V^M 
\end{eqnarray}
From eqs. (\ref{eq:Dirac}-\ref{eq:light}) we see that the
mixing matrix $V^M$ diagonalize the following symmetric matrix:
\begin{eqnarray}\label{eq:C}
{\cal C}&=&M_D^\Delta V_0^\dagger \frac{1}{M_R} V_0^\star M_D^\Delta
\end{eqnarray}
where $V_0$ is the mixing matrix that diagonalize on the right the
Dirac neutrino mass matrix in eq. (\ref{eq:Dirac}).

\subsection{$V^M$ as correlation matrix in GUT}

In GUT models, such us generic $SO(10)$ or $E_6$,
there are some natural Yukawa
unifications. These give up to an interesting relation between the $U^{CKM}$
quark mixing matrix, $U^{PMNS}$ lepton mixing matrix and $V^M$ obtained from
eq. (\ref{eq:C}). In fact $V^M$ turns out to the the correlation matrix defined
in eq. (\ref{eq:fund}).
In the quark sector we introduce $Y_u$ and $Y_d$
to be the Yukawa matrices for up and
down sectors.
They can be diagonalized by
\begin{eqnarray}
Y_u&=&U_u Y_u^\Delta V_u^\dagger\\
Y_d&=&U_d Y_d^\Delta V_d^\dagger
\end{eqnarray}
where the $Y^\Delta$ are diagonal and the $U$s and $V$s are unitary matrices.
\\
Then the quark mixing matrix is given by
\begin{eqnarray}
U^{CKM}&=&U_u^\dagger U_d
\end{eqnarray}
In GUT models such as $SO(10)$ or $E_6$ we have intriguing relations between the
Yukawa coupling of the quark sector and the one of the lepton sector.
For instance, in minimal renormalizable $SO(10)$ with Higgs in the
$\bf 10$, $\bf 126$, and $\bf 120$, we have $Y_l \approx Y_d^t$.
In fact the flavor symmetry implies the structure of the Yukawa matrices:
the equivalent entries of $Y_l$ and $Y_d$ are usually of the same order
of magnitude.
We conclude that, as long as the flavor symmetry fully constraints the
mixing matrices that diagonalize a Yukawa matrices, we have $U_l \simeq V_d^\star$.

From eq. (\ref{eq:PMNS}) we get
$$ U^{PMNS} \simeq V_d^t U_0 V^M$$
where $V^M$ is the mixing matrix that diagonalize the matrix $\cal C$ of eq.
(\ref{eq:C}).
If we call $Y_\nu$ the Yukawa coupling that will generate the Dirac neutrino
mass matrix $M_D$, we have also the relation
\begin{eqnarray}
Y_\nu \approx Y_u^t &\rightarrow& U_0 \simeq V_u^\star
\end{eqnarray}
This relation, together with the previous one, implies
$$U^{PMNS} \simeq V_d^t V_u^\star V^M$$
If the Yukawa matrices are symmetric, for example in minimal renormalizable
$SO(10)$ we have only small contributions from the antisymmetric representation
${\bf 120}$, the previous relationship translates into a relation between
$U^{PMNS}$, $U^{CKM}$ and $V^M$.
In fact we have
\begin{eqnarray}\label{VM}
Y_u   = Y_u^t &\rightarrow& V_u^\star = U_u\\
Y_d   = Y_d^t &\rightarrow& V_d^\star = U_d\,. \nonumber
\end{eqnarray}
The first relation tell us that 
$$U^{PMNS} = V_d^t U_u V^M\,.$$
Finally, by using the second relation in eq. (\ref{VM})
and the definition of the CKM mixing
matrix of eq. (\ref{CKM}) we get that
$$U^{PMNS} = (U^{CKM})^\dagger V^M$$

The form of $V^M$ can be obtained under some assumptions about the
flavor structure of the theory.
This model will give for example a correlation $V^V$ with $V^M_{13}=0$.
As a consequence of the from of quark-lepton correlation matrix $V^M$
there are some predictions for the model. 
For example the prediction for $\theta_{13}^{PMNS}$ of \cite{Chauhan:2006im}
and the correlations between CP violating phases and the mixing angle
$\theta_{12}$ of sect.~{\bf\ref{sec:CP}}.

\subsection{$V^M$ from $S_3$ flavor symmetry}
As we told $V^M$ diagonalize the matrix ${\cal C}$ of eq. (\ref{eq:C}).
In this section we will show that a $S_3$ flavor permutation symmetry,
softly broken into $S_2$, gives us the prediction of $V^M_{13}=0$.

The six generators of the $S_3$ flavor symmetry are
the elements of the permutation group of three objects.
The action of $S_3$ on the fields is to permute the
family label of the fields.
In the following we will introduce the $S_2$ symmetry with respect the 2nd
and 3rd generations. The $S_2$ group is an Abelian one
and swap 
the second family $\{\mu_L, (\nu_\mu)_L, s_L, c_L, \mu_R, (\nu_\mu)_R, s_R, c_R\}$ with
the third one $\{\tau_L, (\nu_\tau)_L, b_L, t_L, \tau_R, (\nu_\tau)_R, b_R, t_R\}$.

Let us assume that there is an $S_3$ flavor symmetry at high energy, which is
softly broken into $S_2$.
In this case, before the $S_3$ breaking all the Yukawa matrices have the
following structure:
\begin{eqnarray}\label{eq:YS3}
Y &=&\pmatrix{ 
a&b&b\cr
b&a&b\cr
b&b&a}
\end{eqnarray}
where $a$ and $b$ independent. The $S_3$ symmetry implies 
that $(1/\sqrt{3},1/\sqrt{3},1/\sqrt{3})$ is an eigenvector
of our matrix in eq. (\ref{eq:YS3}).
Moreover these kind of matrices have two equal eigenvalues.
This gives us an undeterminated mixing angle 
in the diagonalizing mixing matrices.

When $S_3$ is softly broken into $S_2$, one get
\begin{eqnarray}\label{eq:YS2}
Y &=&\pmatrix{ 
a&b&b\cr
b&c&d\cr
b&d&c}
\end{eqnarray}
with $c\approx a$ and $d\approx b$. When $S_3$ is broken the degeneracy
is removed.
In general the $S_2$ symmetry implies that $(0,1/\sqrt{2},-1/\sqrt{2})$
is an exact eigenvector of our matrix (\ref{eq:YS2}).
The fact that $S_3$ is only softly broken
into $S_2$ allows us to say that $(1/\sqrt{3},1/\sqrt{3},1/\sqrt{3})$
is still in a good approximation an eigenvector of Y in eq. (\ref{eq:YS2}).
Then the mixing matrix that diagonalize from the right
the Yukawa mixing matrix in eq. (\ref{eq:YS2}) is given
in good approximation by
\begin{eqnarray}\label{eq:TBM}
\pmatrix{
\approx -\sqrt{2}/\sqrt{3}& \approx 1/\sqrt{3}& 0\cr
\approx        1/\sqrt{6}& \approx 1/\sqrt{3}& 1/\sqrt{2}\cr
\approx        1/\sqrt{6}& \approx 1/\sqrt{3}&-1/\sqrt{2}\cr
}
\end{eqnarray}
where we did not prompt the phases.

Let us now investigate the $V^M$ in this model.
The mass matrix $M_D$ will have the general structure in eq. (\ref{eq:YS2}).
To be more defined,
let us assumed that there is an extra
softly broken $Z_2$ symmetry under which the 1st and the 2nd families are
even, while the 3rd family is odd.
This extra softly broken $Z_2$ symmetry gives us a hierarchy between
the off-diagonal and the diagonal elements of $M_D$, i.e. $b,d<< a,c$.
In fact if $Z_2$ is exact both $b$ and $d$ are zero.
For simplicity, we assume also a quasi-degenerate spectrum for
the eigenvalues of the Dirac neutrino matrix as in \cite{Caravaglios:2006aq}.

The Majorana right handed neutrino is of the form
\begin{eqnarray}
M_R &=&\pmatrix{ 
a&b&b^\prime\cr
b&c&d\cr
b^\prime&d&e}
\end{eqnarray}
Because $S_3$ is only softly broken into $S_2$ we have that
$a\approx c \approx e$, and $b\approx b^\prime\approx d$.
In this approximation the $M_R$ matrix is diagonalized by a $U_R$ of the
form in eq. (\ref{eq:TBM}).
In this case we have that $M_\nu$ defined in eq. (\ref{eq:Mnu})
is near to be $S_3$ and $S_2$ symmetric, then it is diagonalized by a
mixing matrix $U_\nu$ near the TriBiMaximal one given in eq. \ref{eq:TBM}.
The ${\cal C}$ matrix is diagonalized by the mixing matrix
\begin{equation}
V_M=U_\nu U_R
\end{equation}
We obtain that $V_M$ is a rotation in the $(1,2)$ plan,
i.e. it contains a zero in the $(1,3)$ entry.

As shown in \cite{Caravaglios:2006aq}, it is possible to fit the
CKM and the PMNS mixing matrix within this model
\footnote{Notice that in \cite{Caravaglios:2006aq} the mass matrix $M_d$
is not symmetric, however the mixing matrix $V_d^\star$ is similar to
the mixing matrix $U_e$.}.

\section{Summary and conclusions}\label{sec:End}
In this paper we show the power of the quark lepton correlation
in GUT models. In particular we investigated the correlation
between the CP violating invariants and the mixing angle $\theta_{12}$
in the lepton sector. To extract these informations
we used a Monte Carlo approach to take into full
account the presence of unknow unphysical phases in the definition
of $V^M$.

We obtain that
\begin{eqnarray*}
\theta_{13}^{PMNS}&=&(9^{+1}_{-2})^\circ\\
J,S_1,S_2 &&\mbox{ are correlated to }\sin^2\theta_{12}^{PMNS}
\end{eqnarray*}
with the theoretical input $V^M_{13}=0$ only.
The results are such that in the next future it will be
possible to make cross check from the experimental evidences
and discriminate the validity of this approach.

To better clarify the importance of the quark lepton correlation
we shown how a toy model, based on the $S_3$
flavor permutation symmetry and GUT, predicts the correlation matrix
$V^M$ to have a zero $(1,3)$ entry.
In this model $V^M$, defined by \ref{eq:fund}
is also related to the Dirac and Majorana neutrino mass matrix
as in eq. \ref{eq:C}.
For this reason its form is given by the symmetries of the model.
In this model we have $V^M_{13}=0$. We can apply all the nice
predictions obtained in section {\bf\ref{sec:CP}}.

\subsection*{Acknowledgments}
I thank B.C. Chauhan, J. Pulido, and E. Torrente-Lujan
for discussions about neutrino physics. 
I thanks A. Ibarra and S. Morisi for a stimulating
input about neutrino mass in $SO(10)$ unified theories.
I am gratefully to F. Caravaglios for the clarification
of the r\^ole of $S_3$ symmetry in the third quantization \cite{Caravaglios:2002ws}.
I acknowledge the MEC-INFN grant, CYCIT-Ministerio of Educacion (Spain) grant,
and I would like to thank the Physics Department of the University of Milan
for kind hospitality.


\begin{thebibliography}{10}
\bibitem{SNO}
B.~Aharmim {\it et al.}  [SNO Collaboration],
Phys.\ Rev.\ C {\bf 72} (2005) 055502
\bibitem{Krauss:2006qq}
C.~B.~Krauss  [SNO Collaboration],
J.\ Phys.\ Conf.\ Ser.\  {\bf 39} (2006) 275.
\bibitem{SKatm}
[Super-Kamiokande Collaboration],
arXiv:hep-ex/0607059.
\bibitem{SKatm2}
J.~Hosaka {\it et al.}  [Super-Kamiokande Collaboration],
Phys.\ Rev.\ D {\bf 74} (2006) 032002
\bibitem{SKsolar}
J.~Hosaka {\it et al.}  [Super-Kamkiokande Collaboration],
Phys.\ Rev.\ D {\bf 73} (2006) 112001
\bibitem{GNO}
M.~Altmann {\it et al.}  [GNO Collaboration],
Phys.\ Lett.\ B {\bf 616} (2005) 174
%
\bibitem{GALLEX}
W.~Hampel {\it et al.}  [GALLEX Collaboration],
Phys.\ Lett.\ B {\bf 447} (1999) 127.
%
\bibitem{HOMESTAKE} 
B.~T.~Cleveland {\it et al.} [HOMESTAKE],
Astrophys.\ J.\  {\bf 496} (1998) 505.
\bibitem{SAGE}
J.N. Abdurashitov et al. [SAGE Collaboration] Phys. Rev. Lett. 83(23) (1999)4686.
\bibitem{KamLAND} 
T.~Araki {\it et al.}  [KamLAND Collaboration],
Phys.\ Rev.\ Lett.\  {\bf 94} (2005) 081801
\bibitem{CHOOZ}
M.~Apollonio {\it et al.}  [CHOOZ Collaboration],
Phys.\ Lett.\ B {\bf 466} (1999) 415
\bibitem{PaloVerde} 
Y.~F.~Wang  [Palo Verde Collaboration], 
Int.\ J.\ Mod.\ Phys.\ A {\bf 16S1B}, 739 (2001); 
\bibitem{PaloVerde2}
F.~Boehm {\it et al.} [Palo Verde Collaboration], 
Phys.\ Rev.\ D {\bf 64}, 112001 (2001). 
\bibitem{MINOS}
[MINOS Collaboration],
arXiv:hep-ex/0607088.
\bibitem{Charles:2004jd}
  J.~Charles {\it et al.}  [CKMfitter Group],
  Eur.\ Phys.\ J.\ C {\bf 41} (2005) 1
\bibitem{Fogli:2005gs}
G.~L.~Fogli, E.~Lisi, A.~Marrone, A.~Palazzo and A.~M.~Rotunno,
arXiv:hep-ph/0506307.
\bibitem{Aliani:2003ns}
P.~Aliani, V.~Antonelli, M.~Picariello and E.~Torrente-Lujan,
arXiv:hep-ph/0309156.
\bibitem{Aliani:2002na}
P.~Aliani, V.~Antonelli, M.~Picariello and E.~Torrente-Lujan,
Phys.\ Rev.\ D {\bf 69} (2004) 013005
\bibitem{Balantekin:2004hi}
A.~B.~Balantekin, V.~Barger, D.~Marfatia, S.~Pakvasa and H.~Yuksel, 
arXiv:hep-ph/0405019. 
\bibitem{Oberauer:2004ji} 
L.~Oberauer, 
Mod.\ Phys.\ Lett.\ A {\bf 19} (2004) 337 
\bibitem{Rodejohann:2006ek}
  W.~Rodejohann,
  Phys.\ Scripta {\bf T127} (2006) 35.
\bibitem{Gonzalez-Garcia:2006wm}
  M.~C.~Gonzalez-Garcia, M.~Maltoni and J.~Rojo,
  arXiv:astro-ph/0608107.
\bibitem{Giunti:2006fr}
  C.~Giunti,
  arXiv:hep-ph/0608070.
\bibitem{Valle:2006vb}
  J.~W.~F.~Valle,
  arXiv:hep-ph/0608101.
\bibitem{Fogli:2006jk}
  G.~L.~Fogli, E.~Lisi, A.~Mirizzi, D.~Montanino and P.~D.~Serpico,
  arXiv:hep-ph/0608321.
\bibitem{Bandyopadhyay:2006jn}
  A.~Bandyopadhyay, S.~Choubey, S.~Goswami and S.~T.~Petcov,
  arXiv:hep-ph/0608323.
\bibitem{Bilenky:2006sn}
  S.~M.~Bilenky,
  arXiv:hep-ph/0607317.
\bibitem{Messier:2006yg}
  M.~D.~Messier,
  eConf {\bf C060409} (2006) 018
  [arXiv:hep-ex/0606013].
\bibitem{Strumia:2006db}
  A.~Strumia and F.~Vissani,
  arXiv:hep-ph/0606054.
\bibitem{Schwetz:2006dh}
  T.~Schwetz,
  Phys.\ Scripta {\bf T127} (2006) 1
  [arXiv:hep-ph/0606060].
\bibitem{Fukugita:2006rm}
  M.~Fukugita, K.~Ichikawa, M.~Kawasaki and O.~Lahav,
  Phys.\ Rev.\ D {\bf 74} (2006) 027302
  [arXiv:astro-ph/0605362].
\bibitem{Chen:2006rk}
B.~L.~Chen, H.~L.~Ge, C.~Giunti and Q.~Y.~Liu,
  arXiv:hep-ph/0605195.
\bibitem{Robertson:2006pk}
  R.~G.~H.~Robertson,
  Prog.\ Part.\ Nucl.\ Phys.\  {\bf 57} (2006) 90
  [arXiv:nucl-ex/0602005].
\bibitem{Petcov:2006yg}
  S.~T.~Petcov,
  AIP Conf.\ Proc.\  {\bf 805} (2006) 135.
\bibitem{McDonald:2006qf}
  A.~B.~McDonald,
  J.\ Phys.\ Conf.\ Ser.\  {\bf 39} (2006) 211.
\bibitem{Petcov:2006gy}
  S.~T.~Petcov and T.~Schwetz,
  arXiv:hep-ph/0607155.

\bibitem{Barger:2006vy}
  V.~Barger, M.~Dierckxsens, M.~Diwan, P.~Huber, C.~Lewis, D.~Marfatia and B.~Viren,
  Phys.\ Rev.\ D {\bf 74} (2006) 073004
  [arXiv:hep-ph/0607177].

\bibitem{unknown:2006mn}
    [Double Chooz Collaboration],
  arXiv:hep-ex/0606025.

\bibitem{Bernabeu:2006az}
  J.~Bernabeu and C.~Espinoza,
  arXiv:hep-ph/0605132.

\bibitem{McFarland:2006pz}
  K.~S.~McFarland  [MINERvA Collaboration],
  arXiv:physics/0605088.

\bibitem{Savvinov:2006pb}
  N.~Savvinov  [OPERA Collaboration],
  arXiv:hep-ex/0602010.

\bibitem{Decowski:2006zg}
  M.~P.~Decowski  [KamLAND Collaboration],
  Acta Phys.\ Polon.\ B {\bf 37} (2006) 245.

\bibitem{Ochoa-Ricoux:2006qn}
  J.~P.~Ochoa-Ricoux  [MINOS Collaboration],
  Prog.\ Part.\ Nucl.\ Phys.\  {\bf 57} (2006) 147.

\bibitem{Kraus:2006qp}
  C.~Kraus  [SNO+ Collaboration],
  Prog.\ Part.\ Nucl.\ Phys.\  {\bf 57} (2006) 150.

\bibitem{Bhattacharya:2006ri}
  S.~Bhattacharya  [INO Collaboration],
  Prog.\ Part.\ Nucl.\ Phys.\  {\bf 57} (2006) 299.

\bibitem{Ciesielski:2006ie}
  R.~Ciesielski  [OPERA Collaboration],
  Acta Phys.\ Polon.\ B {\bf 37} (2006) 1237.

\bibitem{Vignoli:2006jz}
  C.~Vignoli, D.~Barni, J.~M.~Disdier, D.~Rampoldi and G.~Passardi  [ICARUS
                  Collaboration],
  AIP Conf.\ Proc.\  {\bf 823} (2006) 1643.


\bibitem{Aguilar:2006rm}
  J.~A.~Aguilar {\it et al.}  [ANTARES Collaboration],
  arXiv:astro-ph/0606229.

\bibitem{Bouchta:2006rt}
  A.~Bouchta  [IceCube Collaboration],
  arXiv:astro-ph/0606235.

\bibitem{Ahn:2006zz}
  M.~H.~Ahn {\it et al.}  [K2K Collaboration],
  arXiv:hep-ex/0606032.

\bibitem{Balata:2006ue}
  M.~Balata {\it et al.}  [Borexino Collaboration],
  arXiv:hep-ex/0601035.

\bibitem{Aggouras:2006mm}
  G.~Aggouras {\it et al.}  [NESTOR Collaboration],
  Nucl.\ Phys.\ Proc.\ Suppl.\  {\bf 151} (2006) 279.

\bibitem{Broggini:2006rm}
  C.~Broggini  [LUNA Collaboration],
  Prog.\ Part.\ Nucl.\ Phys.\  {\bf 57} (2006) 343.
\bibitem{Barbieri:1999km}
  R.~Barbieri, L.~J.~Hall, G.~L.~Kane and G.~G.~Ross,
  arXiv:hep-ph/9901228.
\bibitem{Altarelli:2006ri}
  G.~Altarelli,
  arXiv:hep-ph/0611117.
\bibitem{deMedeirosVarzielas:2006fc}
  I.~de Medeiros Varzielas, S.~F.~King and G.~G.~Ross,
  arXiv:hep-ph/0607045.
\bibitem{Morisi:2005fy}
  S.~Morisi and M.~Picariello,
  Int.\ J.\ Theor.\ Phys.\  {\bf 45} (2006) 1267
  [arXiv:hep-ph/0505113].
\bibitem{Caravaglios:2005gw}
  F.~Caravaglios and S.~Morisi,
  arXiv:hep-ph/0503234.
\bibitem{Minakata:2004xt}
  H.~Minakata and A.~Y.~Smirnov,
  Phys.\ Rev.\ D {\bf 70}, 073009 (2004)
\bibitem{Ferrandis:2004mq}
  J.~Ferrandis and S.~Pakvasa,
  Phys.\ Lett.\ B {\bf 603} (2004) 184
\bibitem{Pontecorvo:1967fh}
B.~Pontecorvo,
Sov.\ Phys.\ JETP {\bf 26} (1968) 984
[Zh.\ Eksp.\ Teor.\ Fiz.\  {\bf 53} (1967) 1717].
\bibitem{Maki:1962mu}
Z.~Maki, M.~Nakagawa and S.~Sakata,
Prog.\ Theor.\ Phys.\  {\bf 28} (1962) 870.
\bibitem{Cabibbo:1963yz}
N.~Cabibbo,
Phys.\ Rev.\ Lett.\  {\bf 10} (1963) 531.
\bibitem{Kobayashi:1973fv}
M.~Kobayashi and T.~Maskawa,
Prog.\ Theor.\ Phys.\  {\bf 49} (1973) 652.
\bibitem{Chauhan:2006im}
  B.~C.~Chauhan, M.~Picariello, J.~Pulido and E.~Torrente-Lujan,
{\em Quark-lepton complementarity, neutrino and standard model data predict
  $\theta_{13}^{PMNS} = (9^{+1}_{-2})^\circ$},
  arXiv:hep-ph/0605032.
\bibitem{Smirnov:2006qj}
  A.~Y.~Smirnov,
  arXiv:hep-ph/0604213.
\bibitem{Xing:2005ur}
  Z.~z.~Xing,
  Phys.\ Lett.\ B {\bf 618}, 141 (2005)
\bibitem{Dighe:2006zk}
  A.~Dighe, S.~Goswami and P.~Roy,
  arXiv:hep-ph/0602062.
\bibitem{Raidal:2004iw}
  M.~Raidal,
  Phys.\ Rev.\ Lett.\  {\bf 93} (2004) 161801
\bibitem{Georgi:1979df}
H.~Georgi and C.~Jarlskog,
  Phys.\ Lett.\ B {\bf 86} (1979) 297.
\bibitem{Antusch:2005ca}
S.~Antusch, S.~F.~King and R.~N.~Mohapatra,
Phys.\ Lett.\ B {\bf 618} (2005) 150
\bibitem{Hochmuth:2006xn}
  K.~A.~Hochmuth and W.~Rodejohann,
  arXiv:hep-ph/0607103.
\bibitem{Kang:2005as}
  S.~K.~Kang, C.~S.~Kim and J.~Lee,
  Phys.\ Lett.\ B {\bf 619}, 129 (2005)
\bibitem{Cheung:2005gq}
  K.~Cheung, S.~K.~Kang, C.~S.~Kim and J.~Lee,
  Phys.\ Rev.\ D {\bf 72} (2005) 036003
\bibitem{Ellis:1999my}
  J.~R.~Ellis and S.~Lola,
  Phys.\ Lett.\ B {\bf 458}, 310 (1999)
\bibitem{Antusch:2005gp}
  S.~Antusch, J.~Kersten, M.~Lindner, M.~Ratz and M.~A.~Schmidt,
  JHEP {\bf 0503} (2005) 024
\bibitem{Caravaglios:2006aq}
F.~Caravaglios and S.~Morisi,
  arXiv:hep-ph/0611078.
\bibitem{Wolfenstein:1983yz}
  L.~Wolfenstein,
  Phys.\ Rev.\ Lett.\  {\bf 51} (1983) 1945.
\bibitem{Buras:1994ec}
  A.~J.~Buras, M.~E.~Lautenbacher and G.~Ostermaier,
  Phys.\ Rev.\ D {\bf 50} (1994) 3433
\bibitem{Caravaglios:2002ws}
  F.~Caravaglios,
  arXiv:hep-ph/0211183.
\end{thebibliography}
\end{document}